\documentclass[pre,showpacs,showkeys,preprintnumbers,amsmath,amssymb,superscriptaddress,twocolumn]{revtex4}
\usepackage{graphicx}
\usepackage{amsfonts}
\usepackage{amsmath}

\def\erfc{{\rm erfc}}

\def\MSD{{\mathcal M}}
\def\VACF{{\mathcal V}}

\begin{document}

\title{Evidence of hydrodynamic and subdiffusive \\ motion of tracers in a viscoelastic medium}

\author{Denis~S.~Grebenkov}
 \email{denis.grebenkov@polytechnique.edu}
\affiliation{
Laboratoire de Physique de la Mati\`{e}re Condens\'{e}e (UMR 7643), \\ 
CNRS -- Ecole Polytechnique, 91128 Palaiseau, France}

\author{Mahsa Vahabi}
\affiliation{
Laboratoire de Physique de la Mati\`{e}re Condens\'{e}e (UMR 7643), \\ 
CNRS -- Ecole Polytechnique, 91128 Palaiseau, France}

\author{Elena Bertseva}
\affiliation{
Laboratory of Physics of Complex Matter, Ecole Polytechnique F\'ed\'erale de Lausanne (EPFL), \\
Station 3, CH-1015 Lausanne VD, Switzerland }

\author{L\'aszl\'o Forr\'o}
\affiliation{
Laboratory of Physics of Complex Matter, Ecole Polytechnique F\'ed\'erale de Lausanne (EPFL), \\
Station 3, CH-1015 Lausanne VD, Switzerland }

\author{Sylvia Jeney}
\affiliation{
Laboratory of Physics of Complex Matter, Ecole Polytechnique F\'ed\'erale de Lausanne (EPFL), \\
Station 3, CH-1015 Lausanne VD, Switzerland }

\date{\today}

\begin{abstract}
We propose a theoretical model which relies on the generalized
Langevin equation and may account for various dynamical features of
the thermal motion of organelles, vesicles or macromolecules in
viscoelastic media such as polymer networks.  In particular, we
consider inertial and hydrodynamic effects at short times,
subdiffusive scaling at intermediate times, and eventually optical
trapping at long times.  Simple analytical formulas for the mean
square displacement and velocity auto-correlation function are
derived.  The developed theory is applied to the analysis of
fifty-second long trajectories of micron-sized spherical tracers in
actin gels that were acquired at one microsecond temporal resolution
by using optical tweezers single-particle tracking.  For the first
time, both the subdiffusive scaling and hydrodynamic effects are
observed within a single experiment and accurately described by a
minimal phenomenological model.
\end{abstract}







\maketitle

Single-particle tracking (SPT) techniques allow to survey individual
trajectories of organelles, vesicles, macromolecules or artificial
tracers in order to infer the most detailed information about their
dynamics in complex or viscoelastic media, notably in living cells
\cite{Tolic04,Golding06,Arcizet08,Wilhelm08,Wirtz09,Szymanski09,Metzler09,Jeon11,Bertseva12}.
A reliable solution of this stochastic inverse problem requires the
appropriate choice of a theoretical model that accurately describes
all the relevant features of the dynamics (e.g., viscoelastic
properties of the polymer network or the cytoskeleton, hydrodynamic
interactions, and eventual optical trapping).  In a simple viscous
medium (e.g., water), the random trajectory $X(t)$ of a tracer of mass
$m$ diffusing in a harmonic potential of spring constant $k$ can be
described as a solution of the Langevin equation $m\ddot{X}(t) = F(t)
+ F_S(t) - k X(t)$, in which interactions of a tracer with the
surrounding bath are represented by the thermal force $F(t)$ and the
instantaneous Stokes force $F_S(t) = - \gamma \dot{X}(t)$ with the
viscous drag coefficient $\gamma$ \cite{Coffey}.  Two separate
extensions of the Stokes force have been thoroughly studied.  First,
hydrodynamic interactions of a big tracer with the surrounding fluid
were included by adding the Basset force which for a spherical tracer
of radius $a$ reads as
\cite{Basset}
\begin{equation}
\label{eq:F_Basset}
F_B(t) = - \frac{2}{3}\pi a^3 \rho_{\rm f} \ddot{X} - 6a^2 \sqrt{\pi \rho_{\rm f} \eta} \int\limits_{-\infty}^t dt' \frac{\ddot{X}(t')}{\sqrt{t-t'}} ~,
\end{equation}
where $\rho_{\rm f}$ and $\eta$ are the fluid density and viscosity (note
the Stokes-Einstein relation $\gamma = 6\pi a \eta$).  The influence
of hydrodynamic interactions which appear at short time scales, has
been thoroughly investigated, both theoretically
\cite{Widom71,Hinch75,Clercx92,Mainardi96,Indei12a,Indei12b} and
experimentally \cite{Lukic07,Franosch11,Huang11,Grimm11,Grimm12}.
Second, the viscoelastic properties of the medium were included by
replacing the instantaneous Stokes force by a friction memory kernel
$\gamma_\alpha(t)$ \cite{Mori65,Hess83}
\begin{equation}
\label{eq:F_Stokes}
F_S(t) = - \int\limits_{-\infty}^t dt' \gamma_\alpha(t-t') \dot{X}(t').  
\end{equation}
In particular, a slowly decaying kernel $\gamma_\alpha(t) =
\frac{\gamma_\alpha}{\Gamma(1-\alpha)}~ t^{-\alpha}$ with an exponent
$0 < \alpha < 1$ leads to subdiffusive scaling of the mean square
displacement (MSD) at intermediate (or long) times: $\langle
(X(t)-X(0))^2\rangle \propto t^{\alpha}$, where $\langle \ldots
\rangle$ denotes the ensemble average over random realizations of the
thermal force
\cite{Porra96,Wang99,Lutz01,Pottier03,Vinales06,Desposito09,Grebenkov11a}.
The subdiffusive behavior has been observed experimentally in a
variety of viscoelastic media, notably in living cells
\cite{Tolic04,Golding06,Arcizet08,Wilhelm08,Wirtz09,Szymanski09,Metzler09,Jeon11,Bertseva12}.
Although the above two ``mechanisms'' were separately documented,
neither theoretical, nor experimental investigation including both
mechanisms was reported, mainly because of significantly different
time scales at which both mechanisms emerge.

In this letter, we present optical tweezers experiments in which the
hydrodynamic interactions and subdiffusive scaling manifest
simultaneously, and a theory which accurately describes the
experimental results.  In particular, we derive simple analytical
formulas for the MSD and the velocity auto-correlation function (VACF)
that are used for fitting experimental MSD and VACF.  We show that the
inclusion of hydrodynamic interactions is necessary for accurate
analysis of trajectories of big tracers, especially at short times
(several $\mu$s).

The hydrodynamic interactions and subdiffusive scaling can be combined
by including the Basset force (\ref{eq:F_Basset}) and the Stokes force
(\ref{eq:F_Stokes}) into a Generalized Langevin equation (GLE).  Since
experimental data are available from a starting time (that we set to
$0$), one employs the causality principle to cut the integrals in
Eqs. (\ref{eq:F_Basset}, \ref{eq:F_Stokes}) below $0$.  Applying the
standard technique of forward and inverse Laplace transforms to this
linear GLE, one gets a formal solution as
\cite{Desposito09,Grebenkov11a}
\begin{equation}
X(t) = X_0(t) + \int\limits_0^t dt' ~ G(t-t') F(t') ,
\end{equation}
in which the deterministic term $X_0(t)$ (depending on the initial
conditions) and the stochastic term are both determined by the linear
response function $G(t)$ which is defined through its inverse Laplace
transform
\begin{equation}
\tilde{G}(s) = \bigl[m s^2 + \tilde{\gamma}(s) s + k \bigr]^{-1} ,
\end{equation}
where the effective friction kernel $\tilde{\gamma}(s)$ includes both
the hydrodynamic interactions and subdiffusive scaling:
$\tilde{\gamma}(s) = \frac12 m_{\rm f} s + \gamma_h s^{1/2} +
\gamma_\alpha s^{\alpha-1}$, with $\gamma_h = 6\pi a^2 \sqrt{\rho_{\rm
f} \eta}$ for a purely viscous medium, and $m_{\rm f} = 4\pi a^3
\rho_{\rm f}/3$ being the mass of the fluid displaced by the tracer.
Both coefficients $\gamma_\alpha$ and $\gamma_h$ are related to the
viscosity of the medium.  However, finding this relation for
viscoelastic media would require a microscopic model for hydrodynamic
interactions that is beyond the scope of the letter.

When the thermal force $F(t)$ is Gaussian, its distribution is fully
characterized by its mean, $\langle F(t) \rangle = 0$, and the
covariance which is related to the frictional part through the
fluctuation-dissipation theorem: $\langle F(t)F(t')\rangle = k_B T
\gamma(|t-t'|)$, $k_B$ being the Bolzmann's constant, $T$ the absolute
temperature \cite{Kubo,Zwanzig}.  In that case, all the dynamical
properties are fully determined by the effective kernel $\gamma(t)$
or, equivalently, by $G(t)$.  In particular, the exact formulas for
the MSD and VACF from \cite{Desposito09,Grebenkov11a} can be
approximated as
\begin{eqnarray}
\label{eq:MSD}
\langle [X(t_1)-X(t_2)]^2\rangle & \simeq & 2k_B T ~ G^{(1)}(|t_1-t_2|)  , \\
\label{eq:VACF}
\langle \dot{X}(t_1) \dot{X}(t_2)\rangle & \simeq & k_B T~ g(|t_1-t_2|)  , 
\end{eqnarray}
where $g(t)$ and $G^{(1)}(t)$ are the derivative and the primitive of
$G(t)$, respectively (the influence of correction terms was studied in
\cite{Grebenkov11a}).

The practical use of the explicit relations (\ref{eq:MSD},
\ref{eq:VACF}) requires computing inverse Laplace transform of
$\tilde{G}(s)$.  Except few particular or limiting cases, no
analytical representation of $G(t)$ is suitable for fitting
experimental curves in the presence of the long memory friction kernel
(i.e., with $0 < \alpha < 1$).  To overcome this problem, several
shortcuts have been proposed such as using the Laplace transform of
the experimental MSD \cite{Mason95} or ``bypassing the foray into
Laplace space'' through the Fourier transform \cite{Evans09}.

We propose a simple way to get an accurate explicit representation of
the linear response function $G(t)$ and, consequently, of the related
quantities such as the MSD and VACF.  For this purpose, we approximate
the scaling exponent $\alpha$ by a rational number $p/q$, with integer
$p$ and $q$.  Changing $s$ to a new variable $z = (s\tau)^{1/q}$ (with
$\tau$ being an appropriate time scale) allows one to represent the
non-analytic function $1/\tilde{G}(s)$ of $s$ as a polynomial $P(z)$
of dergee $2q$
\begin{equation*}
\frac{\tau^2}{M\tilde{G}(s)} \equiv P(z) = z^{2q} + \frac{\gamma_h \sqrt{\tau}}{M} z^{3q/2} + \frac{\gamma_\alpha \tau^{2-\alpha}}{M} z^p + \frac{k \tau^2}{M} , 
\end{equation*}
where $M = m + m_{\rm f}/2$ is the effective mass (note that $3q/2$ is
integer by setting $q$ to be even).  For convenience, we choose the
time scale $\tau$ by setting the coefficient in front of $z^p$ to $1$:
$\tau = (M/\gamma_\alpha)^{1/(2-\alpha)}$.  The above polynomial has
$2q$ (complex-valued) roots $z_j$.  In a general situation, all the
roots are distinct (i.e., $z_j \ne z_k$ for $j\ne k$) so that one gets
$\frac{M}{\tau^2} \tilde{G}(s) = \sum\limits_{j=1}^{2q} \frac{A_j}{z -
z_j}$, with the coefficients $A_j = \prod\limits_{k\ne j}^{2q} (z_j -
z_k)^{-1}$, from which the inverse Laplace transform yields
\begin{equation}
\label{eq:Gt_1}
G(t) = \frac{\tau}{M} \sum\limits_{j=1}^{2q} A_j  (t/\tau)^{\frac{1}{q}-1} E_{\frac{1}{q},\frac{1}{q}}\bigl(z_j (t/\tau)^{\frac{1}{q}}\bigr) ,
\end{equation}
where $E_{\alpha,\beta}(z)$ is the Mittag-Leffler function
\cite{Gorenflo02,Haubold11}.  From this expression, one deduces the
analytical formulas for $G^{(1)}(t)$ and $g(t)$ which are related to
the MSD and VACF through Eqs. (\ref{eq:MSD}, \ref{eq:VACF}):
\begin{eqnarray}
\label{eq:Gt1}
G^{(1)}(t) &=& \frac{\tau^2}{M} \sum\limits_{j=1}^{2q} A_j~ (t/\tau)^{\frac{1}{q}} E_{\frac{1}{q},\frac{1}{q}+1}\bigl(z_j (t/\tau)^{\frac{1}{q}}\bigr) , \\
\label{eq:gt}
g(t) &=& \frac{1}{M} \sum\limits_{j=1}^{2q} A_j~ (t/\tau)^{\frac{1}{q} - 2} E_{\frac{1}{q},\frac{1}{q} - 1}\bigl(z_j (t/\tau)^{\frac{1}{q}}\bigr).  
\end{eqnarray}
The integral representation of Mittag-Leffler functions
\cite{Gorenflo02} can be used for a rapid and accurate numerical
computation of $G^{(1)}(t)$ and $g(t)$.

\begin{figure}
\begin{center}
\includegraphics[width=80mm]{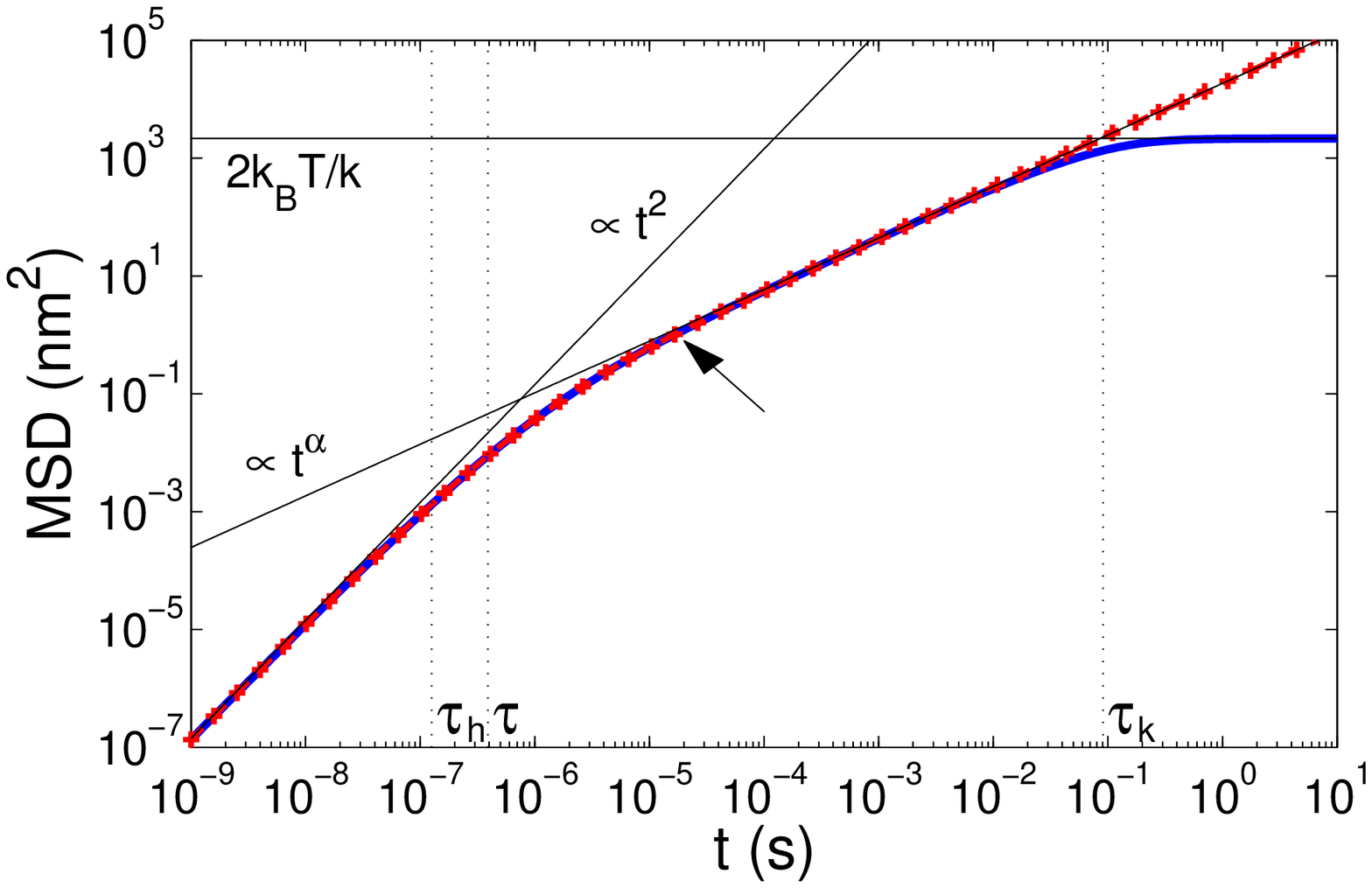}
\includegraphics[width=80mm]{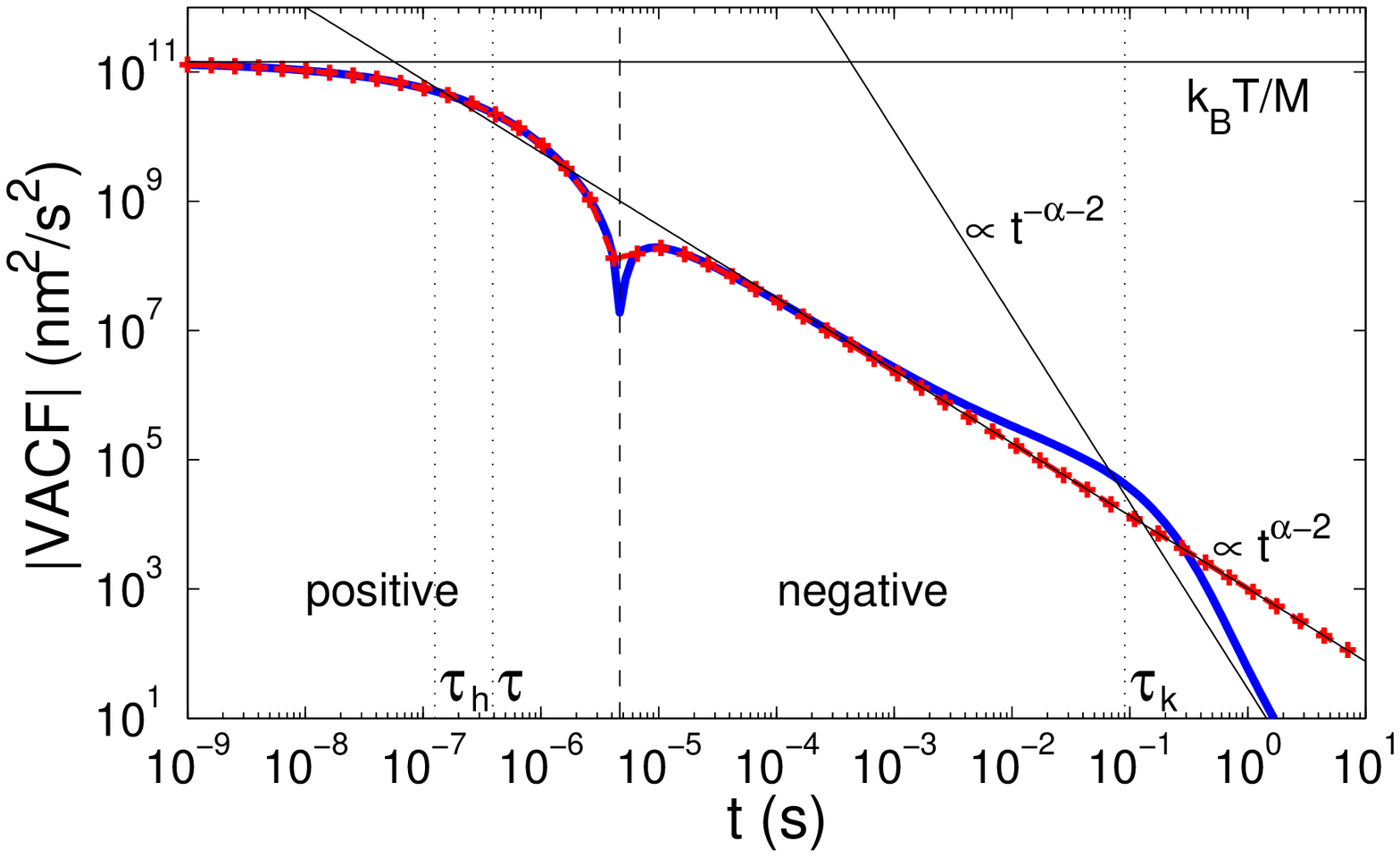}
\end{center}
\caption{
(Color online) Theoretical time-averaged MSD and VACF of a spherical
tracer in a viscoelastic medium.  Curves from Eqs. (\ref{eq:Gt1},
\ref{eq:gt}) are shown by thick solid blue line ($\alpha = 7/8$, $k =
3.77\cdot 10^{-6}$~N/m, $\gamma_\alpha = 4.6\cdot 10^{-7}$
kg/s$^{9/8}$, $\gamma_h = 8.0\cdot 10^{-11}$ kg/s$^{1/2}$, $M =
2.84\cdot 10^{-14}$~kg); curves from Eqs. (\ref{eq:Gt1k0}) without
optical trapping ($k = 0$) are shown by red crosses, and the
asymptotic behaviors at different time scales are shown by thin black
lines.  Vertical dotted lines locate three characteristic time scales
$\tau_h$, $\tau$ and $\tau_k$, while the vertical dashed line
separates the regions with positive and negative values of the VACF.
The arrow indicates the region where the inertial and hydrodynamic
effects lead to deviations from the subdiffusive scaling. }
\label{fig:MSD_theory}
\end{figure}

The above model includes the known explicit solutions:

(i) normal diffusion of a massive tracer ($\alpha = 1$ and $\gamma_h =
0$) corresponds to $p = q = 1$ for which $E_{1,1}(z) = e^{z}$ and thus
$G(t) = (A_1 e^{z_1 \gamma_1 t/M} + A_2 e^{z_2 \gamma_1
t/M})/\gamma_1$, with $z_{1,2} = -\frac12\bigl(1 \pm
\sqrt{1-4kM/\gamma_1^2}\bigr)$.  When $M\to 0$, it reduces to $G(t) =
e^{-kt/\gamma_1}/\gamma_1$.

(ii) normal diffusion of a massive tracer with hydrodynamic
interactions ($\alpha = 1$ and $\gamma_h \ne 0$) is obtained for $p =
q = 2$, for which the functions $E_{\frac12,\frac{n}{2}}(z)$ can be
expressed through $E_{\frac12,1}(z) = e^{z^2} \erfc(-z)$ that yields
the classical formulas for the MSD and VACF
\cite{Clercx92,Lukic07,Grimm12} (here $\erfc(z)$ is the complementary
error function).

(iii) when there is no optical trapping ($k = 0$), some roots $z_j$
are zero so that one gets another representation $\frac{M}{\tau^2}
\tilde{G}(s) = \frac{1}{z^p} \sum\limits_j \frac{\bar{A}_j}{z - z_j}$,
in which $\bar{A}_j = \prod\limits_{k\ne j} (z_j - z_k)^{-1}$ and the
sum and product are taken over all nonzero roots $z_j$ of the
polynomial $P(z)$.  The Laplace inversion of this relation yields
\begin{equation}
G(t) = \frac{\tau}{M} \sum\limits_j \bar{A}_j (t/\tau)^{\frac{1+p}{q}-1} E_{\frac{1}{q},\frac{1+p}{q}}\bigl(z_j (t/\tau)^\frac{1}{q}\bigr),
\end{equation}
from which
\begin{equation}
\label{eq:Gt1k0}
\begin{split}
\hspace*{-2mm} G^{(1)}(t) &= \frac{\tau^2}{M} \sum\limits_j \bar{A}_j (t/\tau)^{\frac{1+p}{q}} 
E_{\frac{1}{q},\frac{1+p}{q}+1}\bigl(z_j (t/\tau)^{\frac{1}{q}}\bigr) , \\
g(t) &= \frac{1}{M} \sum\limits_j \bar{A}_j (t/\tau)^{\frac{1+p}{q} - 2} E_{\frac{1}{q},\frac{1+p}{q} - 1}\bigl(z_j (t/\tau)^{\frac{1}{q}}\bigr).  \\
\end{split}
\end{equation}
The classical formulas for $\alpha = 1$ can be retrieved by setting
again $p = q = 2$ \cite{Widom71,Hinch75}.  In turn, for subdiffusion
with $\gamma_h = 0$, the above formula reduces to $G(t) = \frac{t}{M}
E_{2-\alpha,2}(-(t/\tau)^{2-\alpha})$
\cite{Vinales06,Desposito09,Grebenkov11a}.

Using the properties of Mittag-Leffler functions and relations for the
roots $z_j$ of the polynomial $P(z)$, one can derive the asymptotic
behavior of $G(t)$ (and thus the MSD and VACF) at different time
scales (Fig. \ref{fig:MSD_theory}): 
\begin{equation*}
\begin{split}
G(t) & \simeq \frac{t}{M} - \frac{4\gamma_h t^{\frac32}}{3\sqrt{\pi} M^2}  + \frac{\gamma_h^2 t^2}{2M^3} 
- \frac{8\gamma_h^3 t^{\frac52}}{15\sqrt{\pi} M^4} + \ldots (t\ll \tau_h), \\
G(t) & \simeq \frac{t^{\alpha-1}}{\gamma_\alpha \Gamma(\alpha)} - \frac{\gamma_h t^{2\alpha-\frac52}}{\gamma_\alpha^2 \Gamma(2\alpha-\frac32)} + 
 \ldots  \hskip 9mm (\tau \ll t \ll \tau_k), \\
G(t) & \simeq \frac{\gamma_\alpha t^{-\alpha-1}}{|\Gamma(-\alpha)|k^2} + \frac{\gamma_\alpha^2 t^{-2\alpha-1}}{\Gamma(-2\alpha)k^3}
- \frac{3\gamma_h t^{-\frac52}}{4\sqrt{\pi} k^2} + \ldots  (t \gg \tau_k), \\
\end{split}
\end{equation*}
where $\tau_k = (\gamma_\alpha/k)^{1/\alpha}$ is the trapping time and
$\tau_h = (M/\gamma_h)^2$ is the characteristic time for hydrodynamic
interactions.  For normal diffusion ($\alpha = 1$), one retrieves the
classical behavior: $g(t)\simeq \frac{\gamma_h}{2\sqrt{\pi}
\gamma_1^2}~ t^{-3/2}$ \cite{Widom71} for $k = 0$ and $g(t)\simeq
\frac{15 \gamma_h}{8\sqrt{\pi} k^2}~ t^{-7/2}$ for $k > 0$
\cite{Clercx92}.  


\begin{figure}
\begin{center}
\includegraphics[width=80mm]{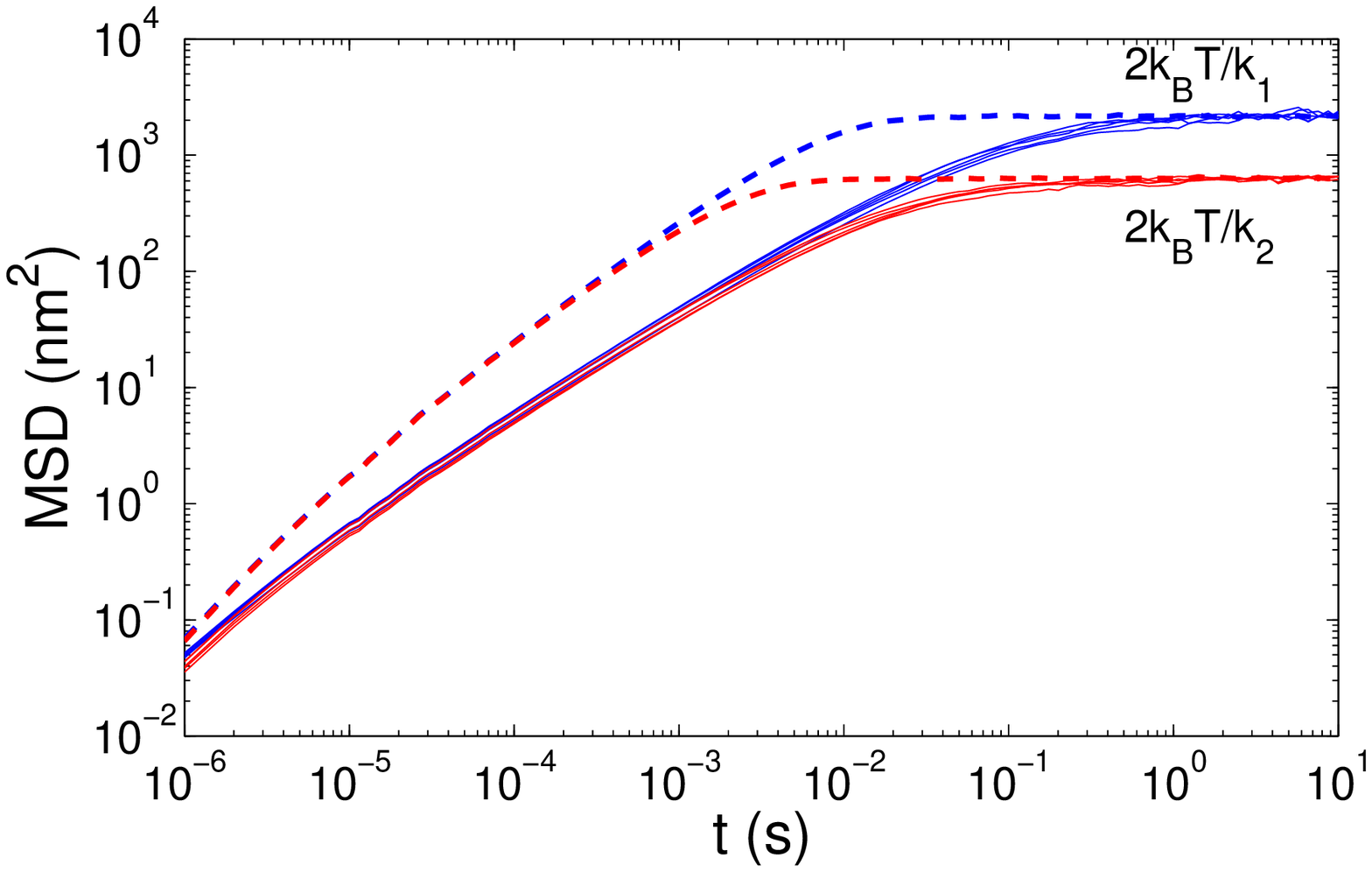}
\includegraphics[width=80mm]{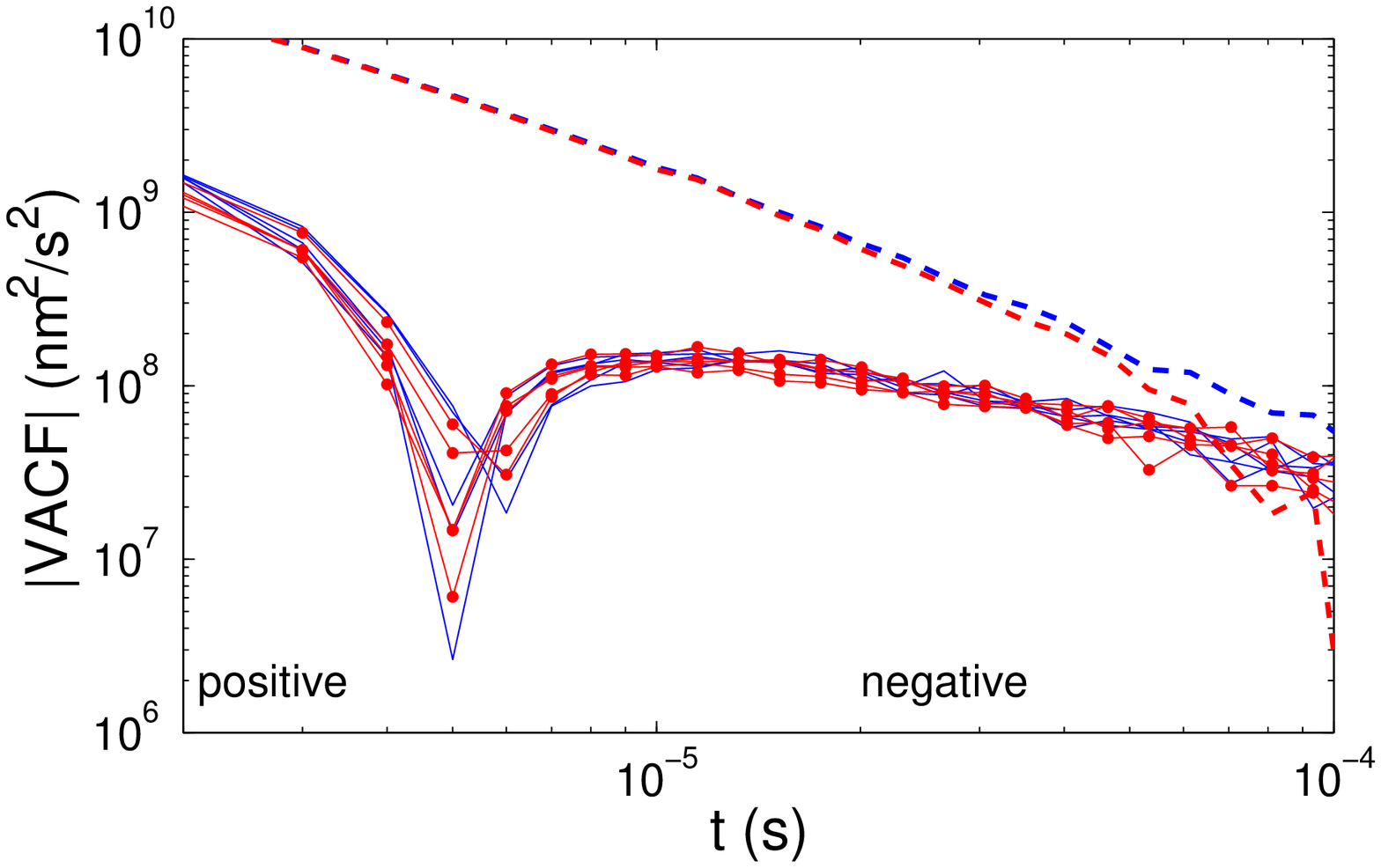}
\end{center}
\caption{
(Color online) Experimental time-averaged MSD and VACF of a spherical
tracer in water (dashed curves) and in an actin gel for the weak
(group of blue solid curves) and strong (group of red solid curves)
optical traps, with trap constants $k_1$ and $k_2$, respectively.
Five curves in each group show the results for five acquisitions.  Red
full circles are added for the VACF plot in order to better
distinguish two groups.  }
\label{fig:MSD_experiment}
\end{figure}


In order to verify the influence of hydrodynamic interactions on
subdiffusion of big tracers, the trajectories of a spherical bead
immersed in an actin gel were recorded.  The experimental set-up has
been described in \cite{Bertseva09,Jeney10}.  The actin filaments were
reconstituted from the pre-formed actin filaments (Cytoskeleton,
Inc. AKF99-A) by dissolving in 2 ml of Milli-Q water.  This gives the
concentration of 0.5 mg/ml in the following buffer: 6 mM Tris-HCl pH
8.0, 0.24 mM CaCl$_2$, 0.24 mM ATP, 2.4 mM MgCl$_2$ and 6\% (w/v)
sucrose.  The 3~$\mu$m diameter melamine resin beads (Microparticles
GmbH, S1712) were added to water beforehand.  The actin filaments with
beads have been placed between a microscope slide and a coverslip
separated by 200 $\mu$m spacers.  The obtained sample chamber was
divided into two parts: one for actin, and the other for pure water.
Both solutions contained the same kind of tracers.

The measurements in water served as a reference for calibration of the
trap.  The time-averaged MSD was computed from the position
fluctuation signal recorded in volts and fitted by $2k_B T \beta^2
G^{(1)}(t)$ with $G^{(1)}(t)$ given by Eq. (\ref{eq:Gt1}) with $p = q
= 2$ (i.e., $\alpha = 1$), $\rho_{\rm f} = 1000$~kg/m$^3$ (water
density), $\rho = 1510$~kg/m$^3$ (melanine resin density), $\eta =
0.959\cdot 10^{-3}$~kg/m/s (dynamic viscosity of water at $T =
295$~K), $a = 1.5~\mu$m (bead radius).  The only fitting parameters
were the spring constant $k$ and the volt-to-meter conversion factor
$\beta$.  For two laser powers, we obtained $k_1 = 3.77\cdot
10^{-6}$~N/m (weak trap) and $k_2 = 1.28\cdot 10^{-5}$~N/m (strong
trap), respectively.  Subsequently, all MSDs recorded in the actin gel
were renormalized to get the same long-time plateau value $2k_B T/k$
of the time-averaged MSD as for beads in water.


After calibration, the time-averaged MSD and VACF are calculated
separately for each bead as
\begin{eqnarray}
\hspace*{-2mm}
\MSD(t) & \equiv &
\frac{1}{N - n} \sum\limits_{k=1}^{N-n} \bigl[X((k+n)\delta) - X(k\delta)\bigr]^2 ,  \\
\VACF(t) & \equiv & \frac{\delta^{-2}}{N-n-1} \hspace*{-2mm} \sum\limits_{k=1}^{N-n-1} \hspace*{-2mm} \bigl[X((k+1)\delta)-X(k\delta)\bigr]   \\
\nonumber
& \times &  \bigl[X((k+n+1)\delta)-X((k+n)\delta)\bigr] , 
\end{eqnarray}
where $X(t)$ is the coordinate of the bead at time $t$, $N$ is the
number of points of the acquired trajectory, $t = \delta n$ is the lag
time, and $\delta$ is the time step of data acquisition.  In the
set-up, $\delta = 1~\mu$s and $N = 5\cdot 10^7$ (fifty-second long
trajectories).

Figure \ref{fig:MSD_experiment} shows the time-averaged MSD and VACF
of a spherical tracer in an actin gel.  The first group of blue solid
curves corresponds to five trajectories in the weak optical trap,
while the second group of red curves corresponds to five trajectories
in the strong optical trap.  The curves from each group are close to
each other.  Deviations between curves within one group can be
attributed to heterogeneities in the actin gel.  Note also that the
curves from both groups are close to each other at small times at
which the optical trap is negligible.  For comparison, the MSD and
VACF for the same bead in water are shown by black dashed lines.

\begin{figure}
\begin{center}
\includegraphics[width=80mm]{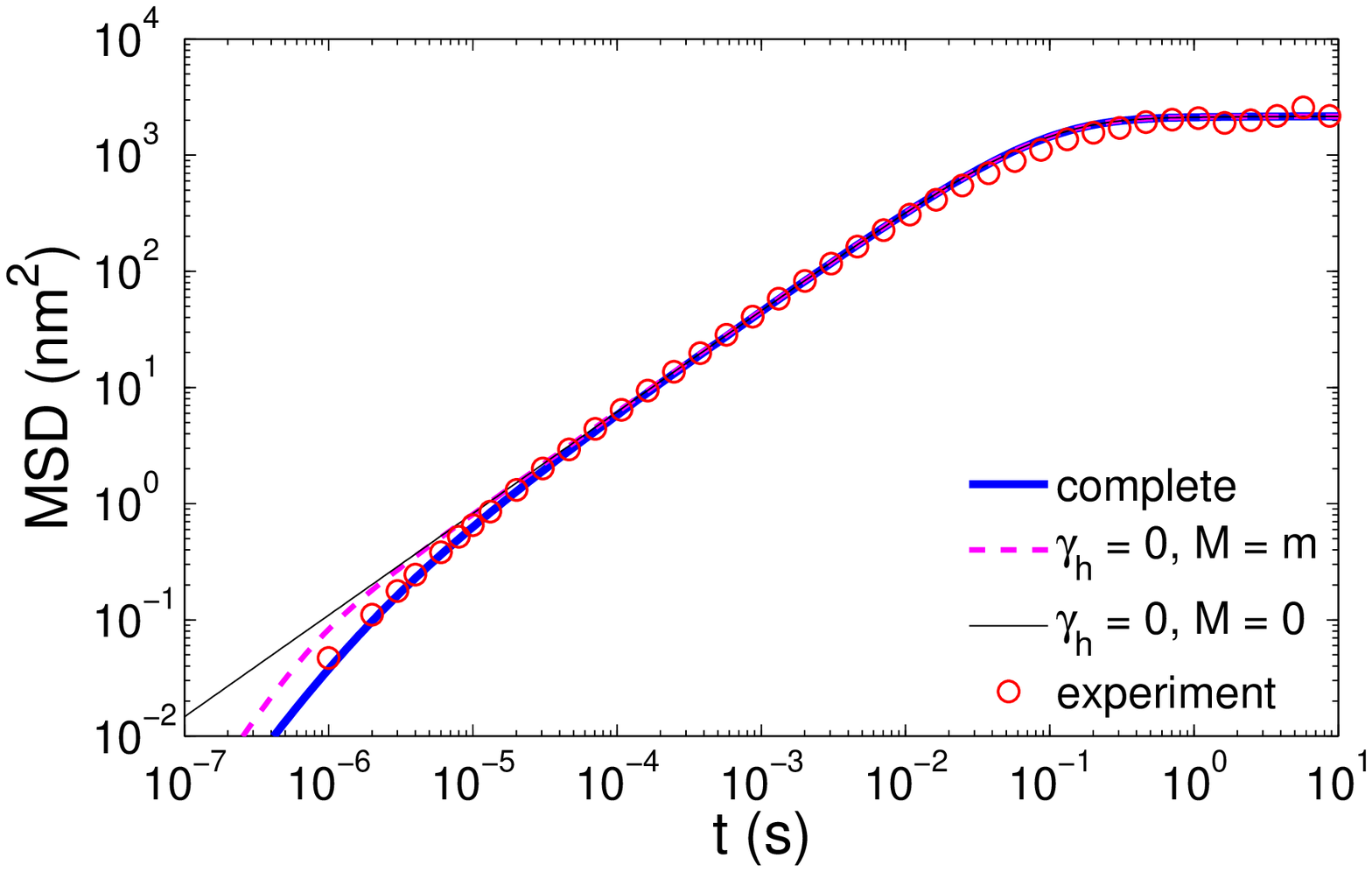}
\includegraphics[width=80mm]{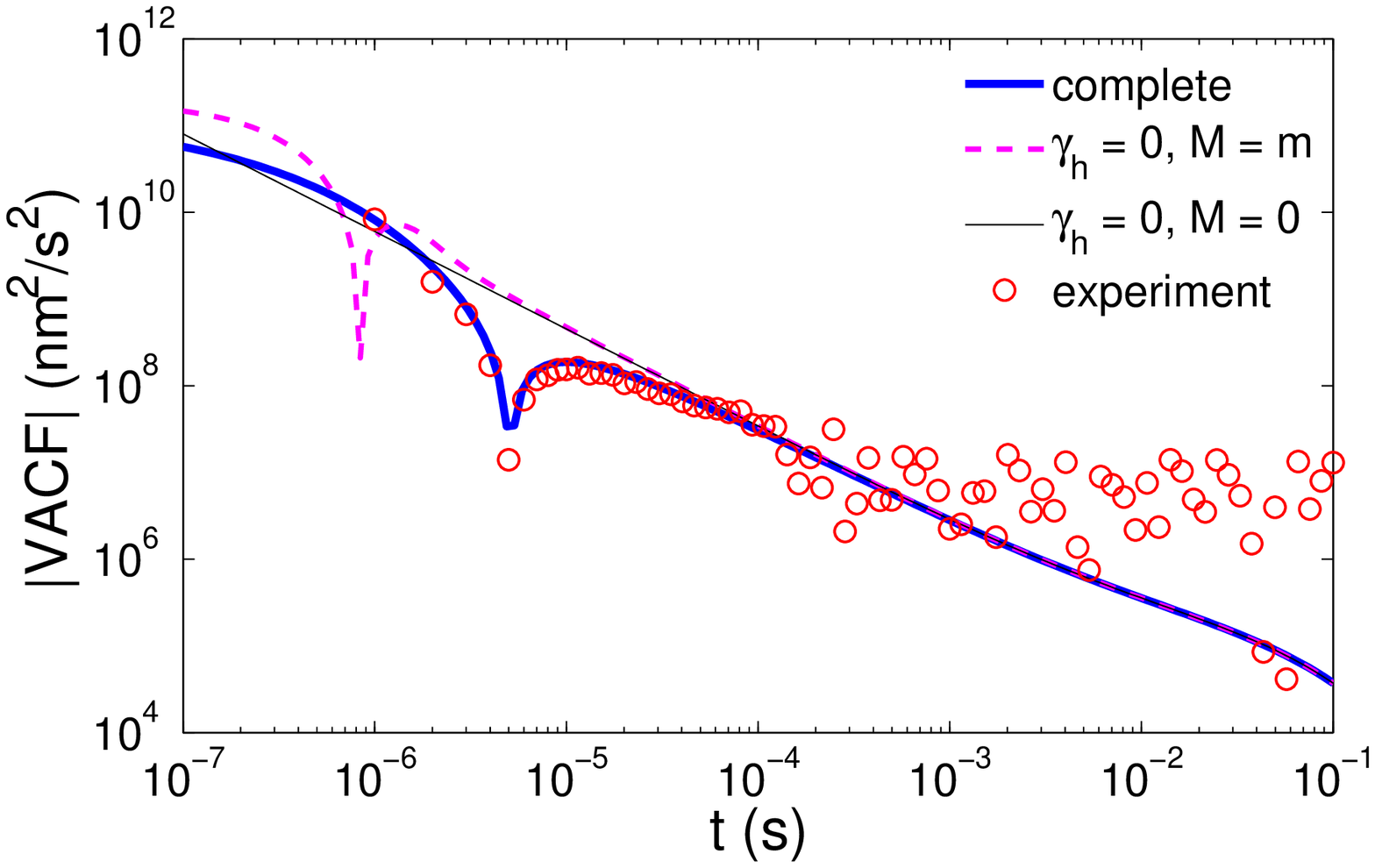}
\end{center}
\caption{
(Color online) Experimental time-averaged MSD and VACF of a spherical
tracer in an actin gel (circles) compared with three theoretical
models: the complete model (anomalous diffusion with hydrodynamic
interactions; shown by solid blue curve), and its two simplified
versions, in which the hydrodynamic interactions and then the inertial
effects are neglected.  The model parameters are: $k = k_1$, $\alpha =
7/8$, $\gamma_\alpha = 4.38\cdot 10^{-7}$ kg/s$^{9/8}$, $\gamma_h =
8.0\cdot 10^{-11}$ kg/s$^{1/2}$, $M = 2.84\cdot 10^{-14}$~kg. }
\label{fig:MSD_models}
\end{figure}

The characteristic feature of big tracers at short times is a clear
deviation from the subdiffusive scaling with the exponent $\alpha$ at
intermediate times (in the range $10^{-5} - 10^{-2}$~s).  The VACF is
particularly sensitive to this deviation which is caused by the
inertial and hydrodynamic effects.  In fact, the MSD and VACF of a
massless tracer are respectively $\frac{2k_B T}{k}(1 -
E_{\alpha,1}(-kt^\alpha/\gamma_\alpha))$ and $\frac{k_B
T}{\gamma_\alpha} t^{\alpha-2}
E_{\alpha,\alpha-1}(-kt^\alpha/\gamma_\alpha)$ that for short times
behave as $\frac{2D_\alpha}{\Gamma(\alpha+1)}~ t^\alpha$ and
$\frac{-D_\alpha}{|\Gamma(\alpha-1)|}~ t^{\alpha-2}$, where $D_\alpha
= k_B T/\gamma_\alpha$ is the generalized diffusion coefficient
\cite{Desposito09,Grebenkov11a}.  In particular, this VACF is always
negative showing anti-persistence in random displacements of a
massless tracer.  In turn, the inertial and hydrodynamic effects
ensure that the VACF becomes positive at short time.  The change of
sign of the VACF at $t = 5-6~\mu$s is clearly seen on
Fig. \ref{fig:MSD_experiment}.

Searching for a minimal phenomenological model to describe the motion
of big spherical tracers in actin gels, one may question whether the
inclusion of the hydrodynamic interactions is relevant or not.  In
other words, would it be sufficient to include only the inertial term
$ms^2$ but neglect the term $\gamma_h s^{3/2}$?  Figure
\ref{fig:MSD_models} illustrates the importance of the hydrodynamic
interactions.  We compare the experimental MSD and VACF (circles) for
one trajectory with the theoretical curves obtained from the complete
model (i.e., when both the inertial and hydrodynamic effects are
included), and two simplified models, in which one first neglects the
hydrodynamic interactions (setting $\gamma_h = 0$ and $M = m$) and
then ignores inertial effects (setting $\gamma_h = 0$ and $M = 0$).
One can see that the theoretical MSD and VACF from Eqs. (\ref{eq:Gt1},
\ref{eq:gt}) (referred to as complete model) correctly reproduce the
experimental ones over 7 orders of magnitude in time (the range being
narrower for the VACF, for which the experimental values become hidden
by intrinsic fluctuations at $t > 10^{-4}$~s).  Deviations between
complete and simplified models for the MSD are relatively small though
visible for the lag time above $1~\mu$s (the range accessible in our
experiment).  In turn, the deviations are much more explicit for the
VACF.  We conclude that the hydrodynamic interactions become
significant for big tracers (of radius above one micron) and at time
scales below tens of microseconds.  The explicit formulas
(\ref{eq:Gt1}, \ref{eq:gt}) allow one to reveal the respective roles
of various forces and mechanisms and to choose the appropriate model
for a reliable analysis of SPT trajectories of endogeneous or
artificial tracers in complex viscoelastic media such as polymer
networks and living cells.

In conclusion, we presented a minimal phenomenological model which
accounts for inertial and hydrodynamic effects at short times,
subdiffusive scaling at intermediate times, and optical trapping at
long times.  All these features have to be included in order to get
accurate fits to the experimental time-averaged MSD and VACF of big
tracers in actin gels when the lag time $t$ ranges from one
microsecond (the acquisition time step in our setup) to tens of
seconds (the measurement duration).  We emphasize on the
phenomenological character of the model in which intricate
interactions of a tracer with actin filaments network are effectively
characterized by the scaling exponent $\alpha$ and the coefficients
$\gamma_\alpha$ and $\gamma_h$.  Microscopic models and/or molecular
dynamics simulations are needed to relate these phenomenological
parameters to mechanical properties of the actin network.  The
advantage of the phenomenological model is that the derived analytical
formulas are explicit and easy to compute through integral
representations of the Mittag-Leffler functions, while the model can
be further extended to incorporate other mechanisms (e.g., adding the
term $s \gamma_1$ to $1/\tilde{G}(s)$ to treat separately the
contributions from the solvent, $\gamma_1$, and from the polymer
networks, $\gamma_\alpha$).

\end{document}